\documentclass{appolb}
\usepackage{graphicx}
\usepackage{cite}

\begin{document}

\title{"Studies of systematic uncertainties of polarization estimation
 for experiments with WASA detector at COSY"%
\thanks{Presented at 
Symposium on applied nuclear physics
and innovative technologies }%
}
\author{M.Hodana, P.Moskal, I.Ozerianska
\address{Jagiellonian University, Cracow\\
IKP-1, Forschungszentrum Juelich, Germany}
\\
{
}
\address{}
}
\maketitle
\begin{abstract}
In November 2010, the azimuthally symmetric WASA detector and the polarized proton beam of
COSY, have been used to collect a high statistics
sample of  $\vec{p}p \rightarrow pp\eta$ reactions in order to determine the
analyzing power as a function of the invariant mass spectra of the two particle subsystems. Here, we show studies
of the influence of the beam and target characteristics such as location and direction on the determination
of the polarization.

\end{abstract}
\PACS{13.88.+e  24.70.+s}
  
\section{Introduction}
In the last decade a vast set of unpolarized cross sections has been determined for the $\eta$ production in the collision of nucleons \cite{Abdel-Bary2003,Moskal2004,Petren2010,Moskal2010,Calen1996,Calen1997,Hibou1998,Smyrski2000,Bergdolt1993,Moskal2009,Moskal2004b,AbdEl2001}. However, the understanding of the production mechanism of this meson still requires the determination of spin observables.
Up to now there are only three measurements of the analyzing power for the $\vec{p}p \rightarrow pp\eta$ 
reaction which have been performed with low statistics and the determined value of the analyzing power is 
essentially consistent with zero \cite{Czyz2007,Winter2003,Balestra2004} within large error bars of 
about $\pm 0.15$.
WASA detector installed at the Cooler Synchrotron COSY gives a possibility to measure the analyzing power with high statistics and high acceptance. Therefore, in November
2010 we have conducted an exclusive measurement of the
$\vec{p}p \rightarrow pp\eta$ reaction using the polarized proton beam of the COSY synchrotron and the WASA detector \cite{MoskalHodana2010}. The
measurement was performed for two beam momenta corresponding to $15$~MeV and $72$~MeV excess energies. The choice of these values of excess energies was dictated by
the availability of the data for the spin averaged cross sections obtained previously
at COSY-11 \cite{Moskal2004}, TOF \cite{Abdel-Bary2003} and WASA/CELSIUS \cite{Petren2010} experiments. 

For the purpose of the monitoring of the degree of  polarization, concurrently to the $\vec{p}p \rightarrow pp\eta$ reaction,
 a proton-proton elastic scattering reactions have been measured. In this contribution we present an estimation  of systematic uncertainties of the determination of the degree of polarization of the COSY beam based on the elastically scattered protons measured by means of the WASA detector setup.

\section{Polarization}
The polarization is extracted using the following formula:
\begin{equation}
P(\theta) = \frac{1}{A_{y}(\theta)\cdot cos\phi}\cdot\frac{N(\theta,\phi) - N(\theta,\phi+\pi)}{N(\theta,\phi) + N(\theta,\phi+\pi)},
\label{e:Pol}
\end{equation}
were $\theta$ is the scattering angle of the forward going proton calculated in the centre of mass frame, $\phi$
is its azimuthal angle, $N$ denotes the number of events and $A_{y}(\theta)$ is the analyzing power of the $\vec{p}p \rightarrow pp$
reaction which was extracted from the results of the EDDA collaboration \cite{Altmeier2000}.

\begin{figure}[htb]
\centerline{%
\includegraphics[width=0.47\linewidth]{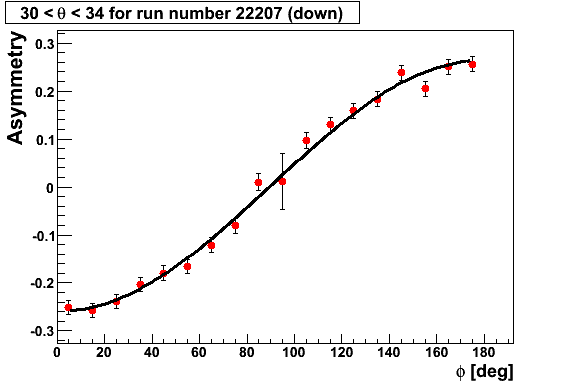}
\includegraphics[width=0.47\linewidth]{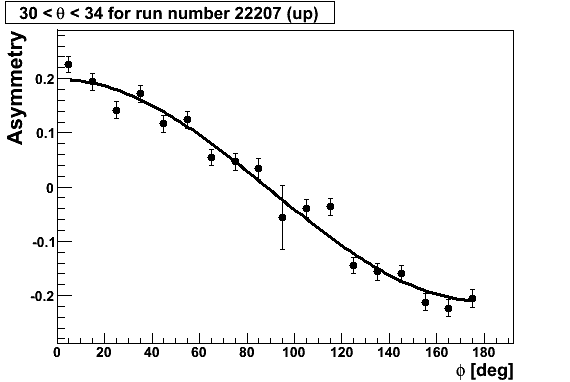}
}
\centerline{%
\includegraphics[width=0.47\linewidth]{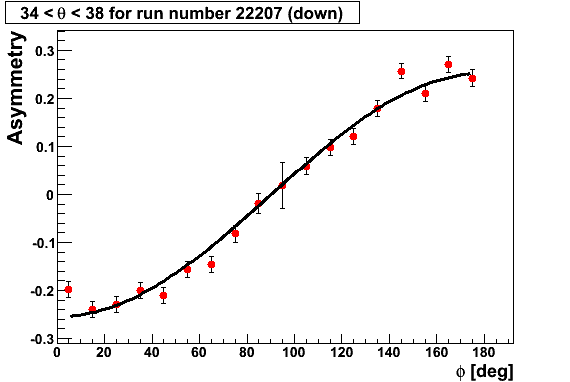}
\includegraphics[width=0.47\linewidth]{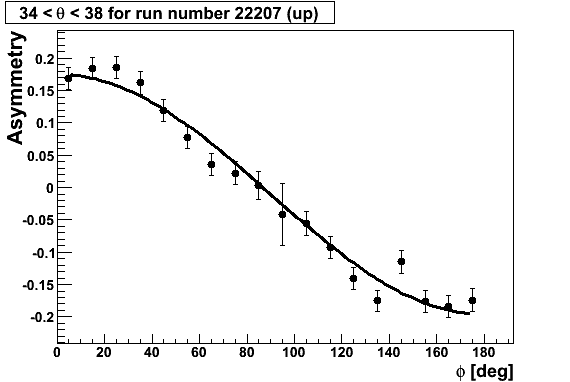}
}
\centerline{%
\includegraphics[width=0.47\linewidth]{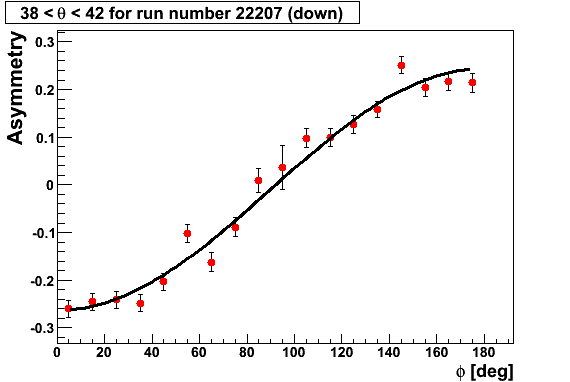}
\includegraphics[width=0.47\linewidth]{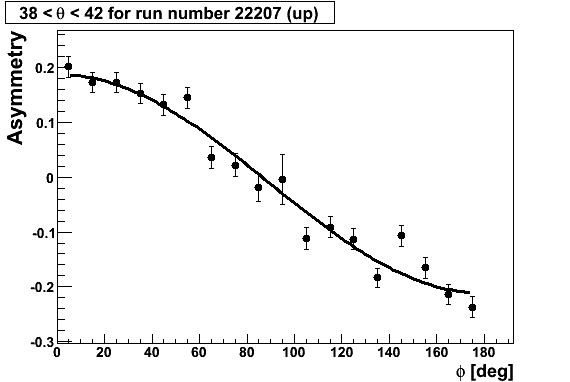}
}
\centerline{%
\includegraphics[width=0.47\linewidth]{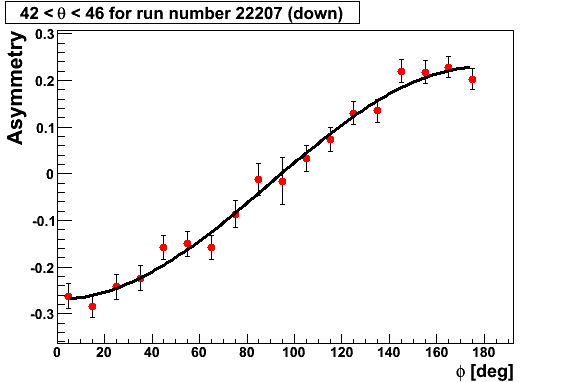}
\includegraphics[width=0.47\linewidth]{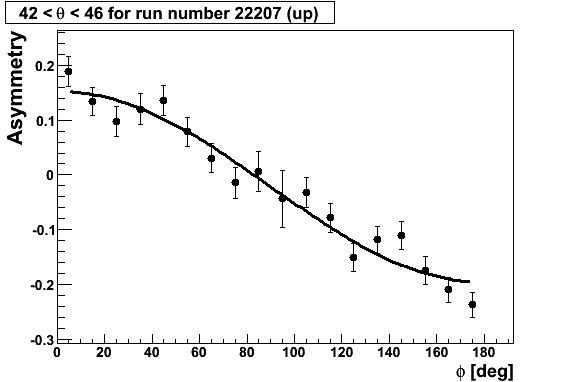}
}
\caption{Experimental distributions of the asymmetry as a function of the proton's azimuthal angle, made for the protons
 scattered into the angle given in histograms' title boxes. The black line represents the fit function given by Eq. \ref{e:P_fit}.
 Left panel: protons with spin down. Right panel: protons with spin up. }
\label{Fig:asymmetryIO}
\end{figure}

The asymmetry, $\epsilon(\theta,\phi)$, is defined as
\begin{equation}
\epsilon(\theta,\phi) =\frac{N(\theta,\phi) - N(\theta,\phi+\pi)}{N(\theta,\phi) + N(\theta,\phi+\pi)}
\label{e:Asym}
\end{equation}
and, according to Eq. \ref{e:Pol}, it can be written as 
\begin{equation}
\epsilon(\theta,\phi) = p_{0}\cdot cos( \phi ),
\label{e:P_fit}
\end{equation}
where $p_{0} = P(\theta) \cdot A_{y}(\theta)$. Polarization is, therefore, extracted by fitting of the function given 
by Eq. \ref{e:P_fit} to $\epsilon(\theta,\phi)$ distributions as shown in Fig.~\ref{Fig:asymmetryIO}. 

The asymmetry is calculated separately for each spin orientation of the polarized protons in four ranges of protons' scattering angle starting from $30^\circ$ up to $46^\circ$ in steps of $4^\circ$.
As a result, four polarizations are extracted for four ranges of the center-of-mass polar angle of the forward scattered proton, $\theta_{CMs}$. The final polarization for a given spin is then calculated as a weighted mean
\begin{equation}
P = \frac{\sum_{i=1}^n{P(\theta_{i})/\sigma^{2}_{P(\theta_{i})}}}{\sum_{i=1}^n 1/\sigma^{2}_{P(\theta_{i})}},
\label{e:P1}
\end{equation}
where $\theta_{i}$ is the scattering angle of the forward going proton, calculated in the centre of mass
system. 

\subsection{Position of the vertex}
The probable source of the systematic uncertainty in the determination of the polarization might be
wrong number of events in the individual $\theta_{CMs}$ ranges, originating from the possible misalignment
of the beam and/or target position.

The reconstruction of tracks of particles registered in the Mini Drift Chamber is free of any
assumption of the position of the reaction vertex. In this respect, obtained angular information 
can be assumed to reflect the actual situation of particles going through the Mini Drift Chamber.
However, reconstruction of tracks of particles going in the forward direction, is based on the assumption that the interaction
point is located at $(x_{v},y_{v},z_{v}) = (0,0,0)$. This may contribute to a systematic uncertainty
of the polarization. To determine the size of this contribution, studies on the position of the interaction
point have been performed.

\begin{figure}[b]
 \begin{minipage}[t]{.55\linewidth}\vspace{0pt}
   \includegraphics[width=\linewidth]{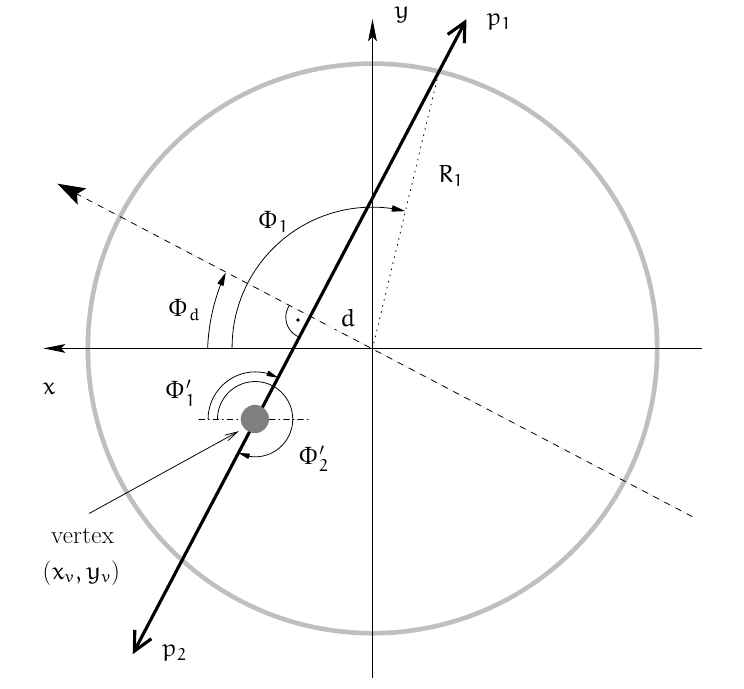}
 \end{minipage}
 \hfill
 \begin{minipage}[t]{.45\linewidth}\vspace{0pt}\raggedright
  \includegraphics[width=\linewidth]{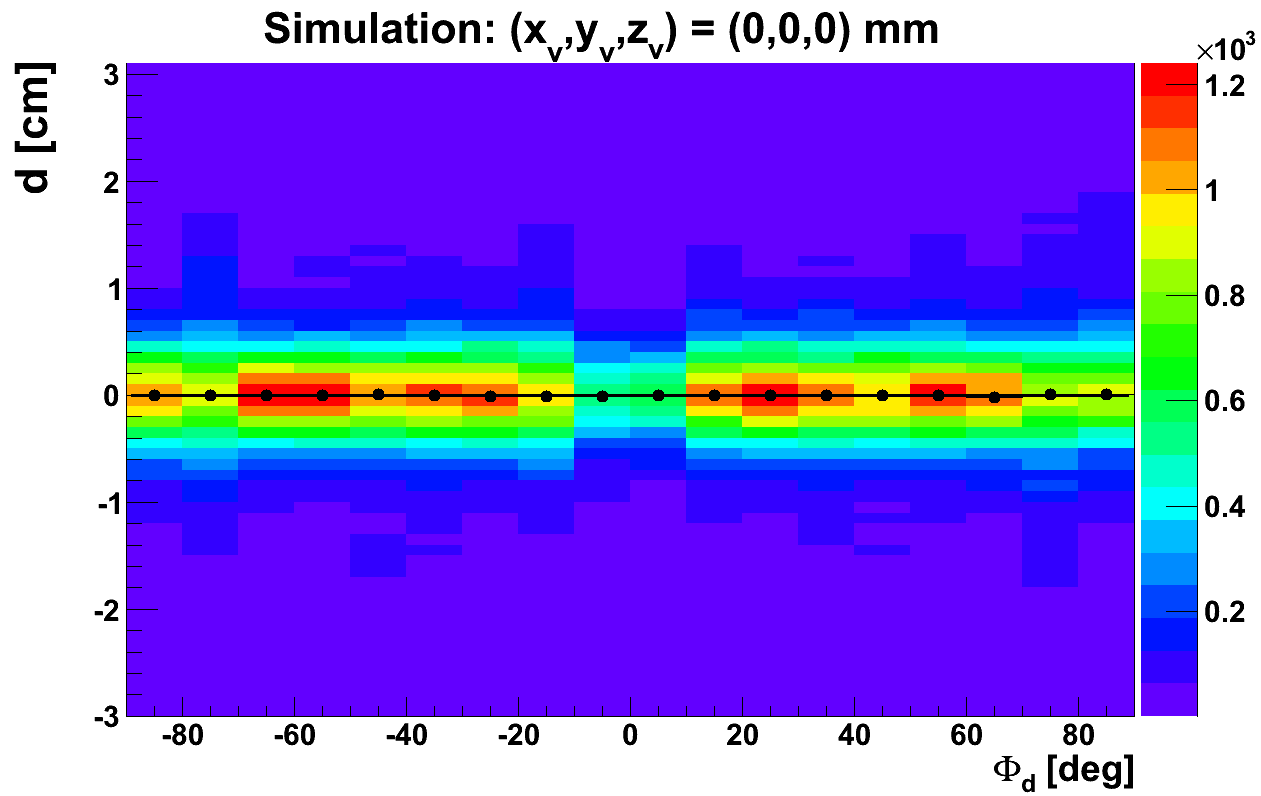}
  \includegraphics[width=\linewidth]{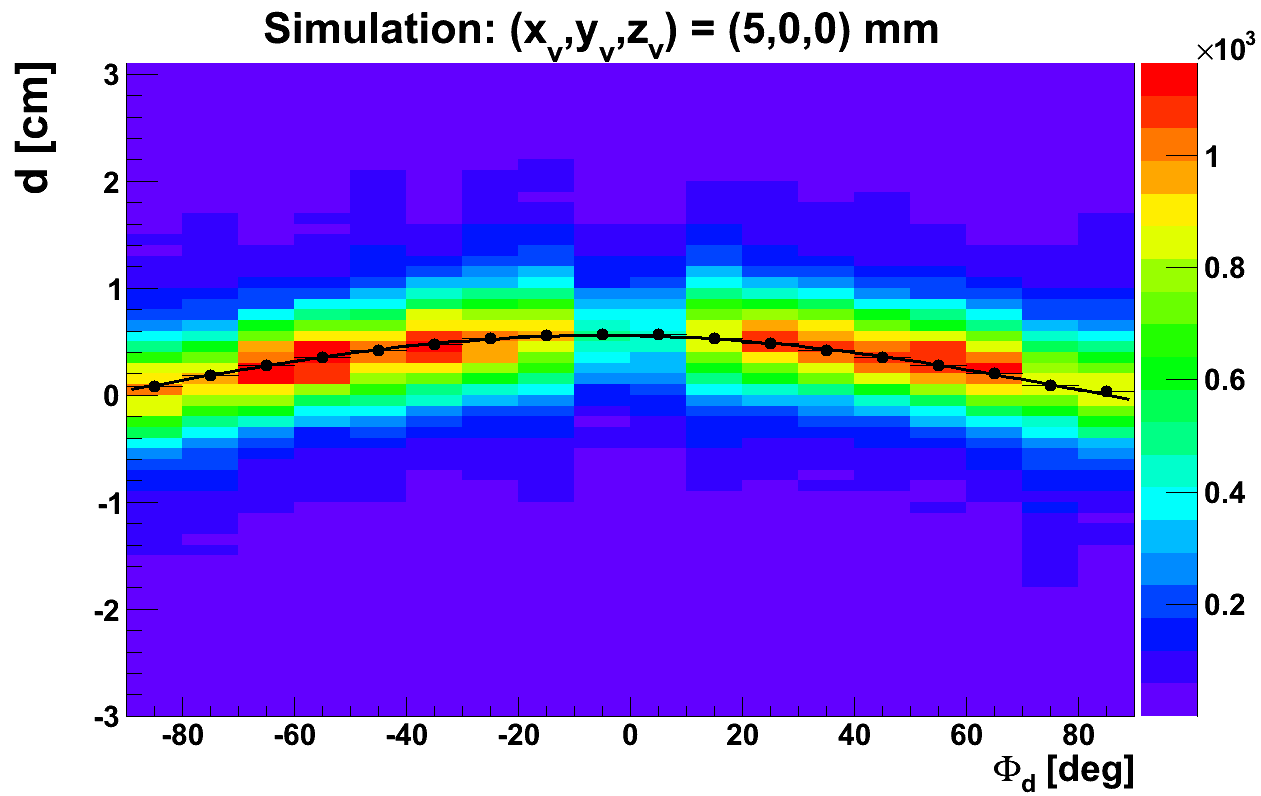}
 \end{minipage}
\caption{Left: picture illustrating the concept of the extraction of the $x_{v}$ and $y_{v}$ coordinates 
of the reaction vertex. Adopted from \cite{Demirors2005}. Right: simulated distributions of $d(\phi_{d})$
made for the vertex position $(x_{v},y_{v},z_{v}) = (0,0,0)$ (upper plot) and $(x_{v},y_{v},z_{v}) = (5,0,0)$~mm 
(lower plot). The points show the positions of the mean of the $d$-distributions for given ranges of $\phi_d$. The line shows a result of the fit of a function given by Eq. \ref{eq:d2}.}
\label{f:ld_xy}
\end{figure}

Fig.~\ref{f:ld_xy} (Left) depicts trajectories of two protons $p_{1}$ and $p_{2}$ projected onto the $(x,y)$ plane.

The~$p_{2}$ proton corresponds to the proton registered in the Mini Drift Chamber. Its reconstructed 
azimuthal angle, $\phi_{2}$, is therefore obtained independently of the position of the reaction vertex,
always reflecting the 'true' value of the emission angle ($\phi_{2} = \phi^{'}_{2}$).
The~$p_{1}$ proton is going in the forward direction
and it intersects the first plane of the Forward Trigger Hodoscope (FTH) at a radius of 
\begin{equation}
R_{1} = Z_{FTH}\cdot tan(\theta_{p_{1}}),
\end{equation}
where $Z_{FTH}$ is the distance from the vertex to the Forward Trigger Hodoscope.
The reconstruction of the path of the $p_{1}$ proton is based on the assumption that the interaction
point is located at $(x_{v},y_{v},z_{v}) = (0,0,0)$. 
Therefore, the reconstructed azimuthal angle $\phi_{1}$, differs from 
the real one $\phi_{1}'$. This disagreement causes deviation from the coplanarity corresponding to 
$\phi_{2}' - \phi_{1}$.

To determine the shift of the reaction vertex, new variables $d$ and $\phi_{d}$ are introduced,
where $d$ is the distance between the point $(0,0,0)$  and the intersection point of dashed line and the solid line in Fig.~\ref{f:ld_xy}. 
The dashed line includes point $(0,0)$ and is perpendicular to the projection of the protons' trajectories.
$\phi_{d}$ is the azimuthal angle between the dashed line and the $x$~-axis.   

With the use of the introduced $d$ and $\phi_{d}$ variables, the $x_{v}$ and $y_{v}$ coordinates 
of the reaction vertex became two parameters in the following formula:
\begin{equation}
d(\phi_{d}) = x_{v}\cdot cos(\phi_{d}) + y_{v}\cdot sin(\phi_{d}).
\label{eq:d2}
\end{equation}
Thus, $x_{v}$ and $y_{v}$ can be extracted by fit of the above function to the $d(\phi_{d})$ distribution
 as shown on the right side of Fig.~\ref{f:ld_xy} for two cases of a vertex
location at $(x_{v},y_{v},z_{v}) = (0,0,0)$ (upper plot) and at $(x_{v},y_{v},z_{v}) = (5,0,0)$~mm 
(lower plot).

\begin{figure}[h]
 \begin{minipage}[t]{.55\linewidth}\vspace{0pt}
   \includegraphics[width=\linewidth]{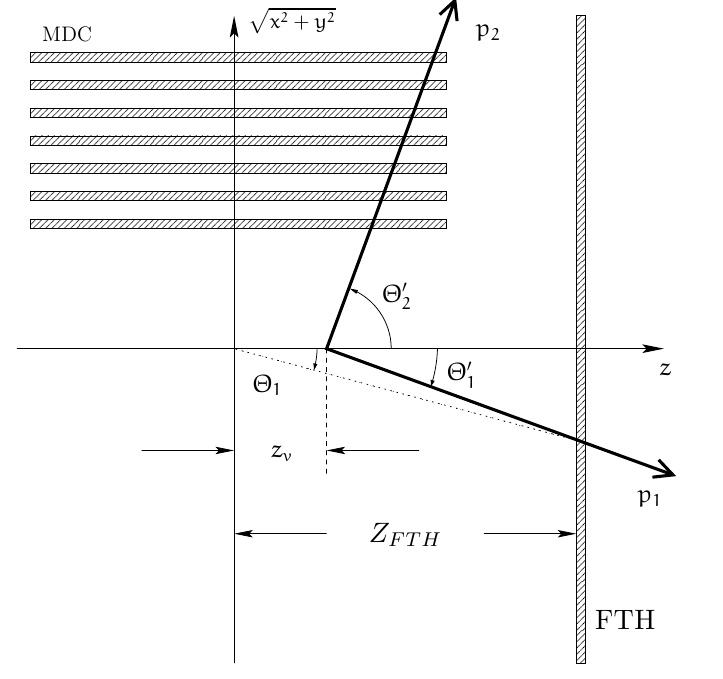}
 \end{minipage}
 \hfill
 \begin{minipage}[t]{.45\linewidth}\vspace{0pt}\raggedright
  \includegraphics[width=\linewidth]{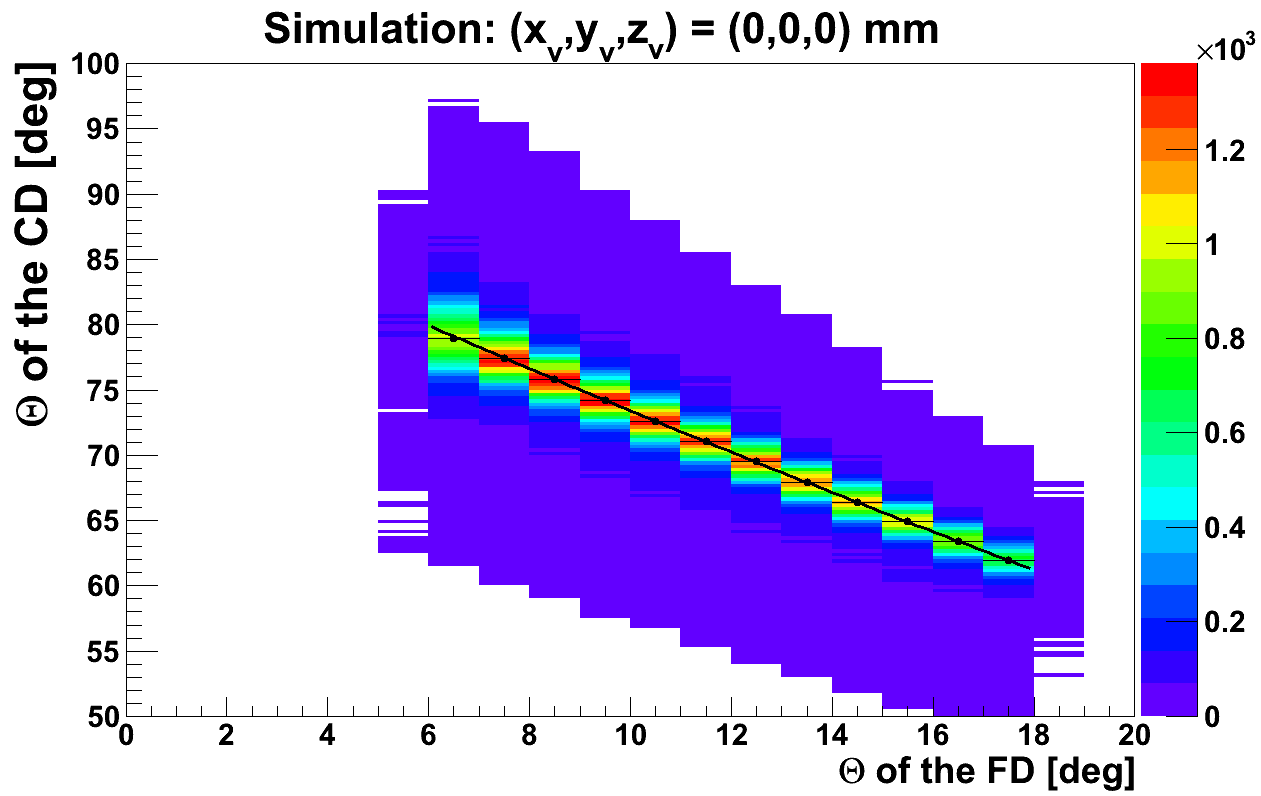}
  \includegraphics[width=\linewidth]{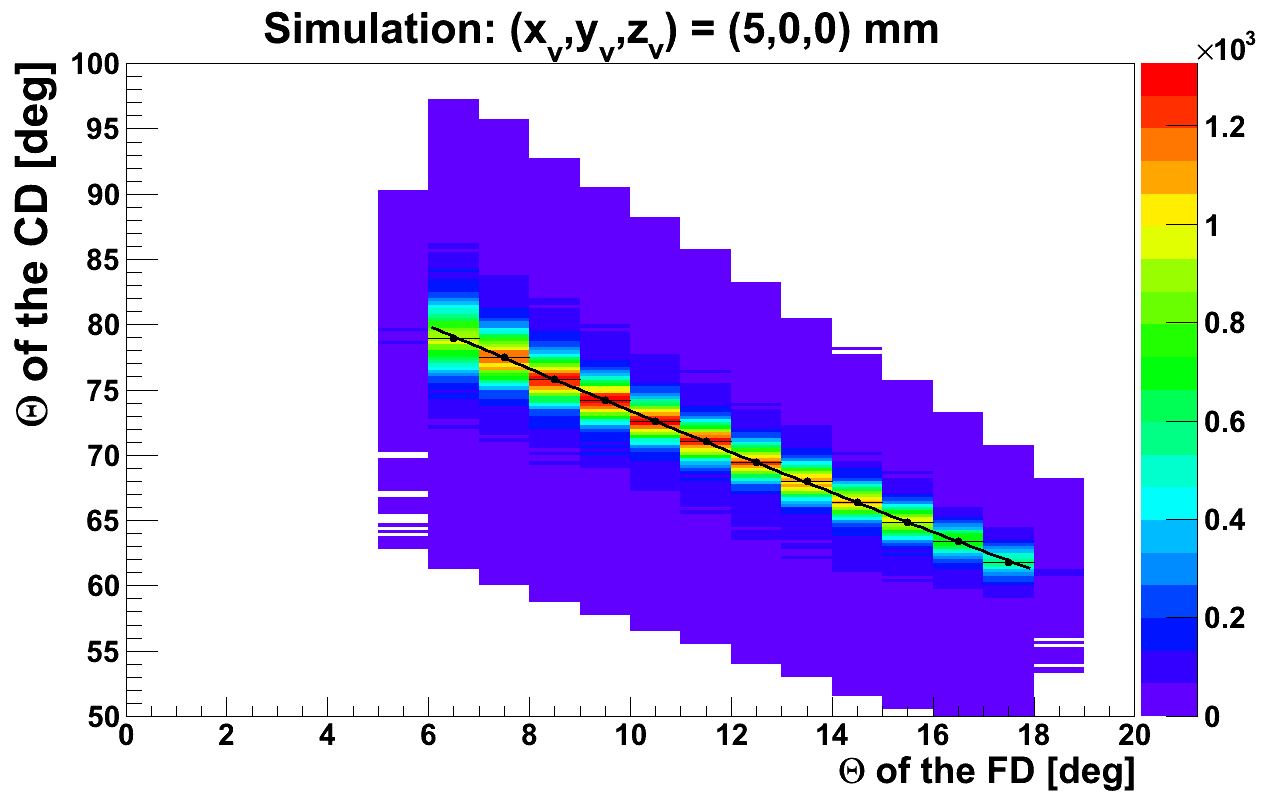}
 \end{minipage}
\caption{Left: picture illustrating the concept of the extraction of the $z_{v}$ coordinate
of the reaction vertex. Adopted from \cite{Demirors2005}. Right: simulated distributions of $\theta_{2}'(\theta_{1})$
made for a vertex position $(x_{v},y_{v},z_{v}) = (0,0,0)$ (upper plot) and $(x_{v},y_{v},z_{v}) = (5,0,0)$~mm 
(lower plot).  The points show the positions of the mean of the $\theta_{CD}$ distribution for given ranges of $\theta_{FD}$. The line denotes result of the fit of  a function given by Eq. \ref{eq:zet} to these points.}
\label{f:ld_z}
\end{figure}

Fig.~\ref{f:ld_z} (Left) depicts the angular dependencies between the two protons $p_{1}$ and $p_{2}$,
used to determine the $z_{v}$ coordinate of the reaction vertex. On the picture, the reaction
vertex is placed on the $z$-axis at the position of $z_{v} > 0 $. 


The trajectory of proton $p_{2}$, reconstructed in the planes of the Mini Drift Chamber, is traced back to the 
actual reaction vertex whereas the track of the forward going proton, $p_{1}$, is 
assumed to origin from the $(0,0,0)$ point. Therefore, the scattering angle $\theta_{1}$ of the forward going
proton deviates from the real value, $\theta_{1}'$.
The relation between the true and reconstructed values of the scattering angle of the
forward going proton can be written as
\begin{equation}
\frac{1}{tan(\theta_{1}')} = \frac{1}{tan(\theta_{1})}(1-\frac{z_{v}}{Z_{FTH}}).
\label{eq:th1}
\end{equation}
Additionally, in an elastic collision the kinematic relation between scattering angles
\begin{equation}
tan(\theta_{1}) \cdot tan(\theta_{2}) = \frac{2\,m_{p}}{2\,m_{p}+T}
\label{eq:kin}
\end{equation} 
must be satisfied, where $m_{p}$ stands for the proton mass and $T$ is the kinetic energy of
the proton beam.

Solving equations \ref{eq:th1} and \ref{eq:kin} for $tan(\theta_{2}')$ results in
\begin{equation}
tan(\theta_{2}')=\frac{ 1-\frac{z_{v}}{Z_{FTH}} }{ tan(\theta_{1})(1+\frac{T}{2\,m_{p}}) }.
\label{eq:zet}
\end{equation}
Thus, the $z_{v}$ coordinate can be extracted by fitting the $\theta_{2}'(\theta_{1})$ distribution.
This is shown on the right side of Fig.~\ref{f:ld_z} for two cases of vertex
location, at $(x_{v},y_{v},z_{v}) = (0,0,0)$ (upper plot) and at $(x_{v},y_{v},z_{v}) = (5,0,0)$~mm 
(lower plot).

\begin{figure}[htb]
\centerline{%
\includegraphics[width=0.32\textwidth]{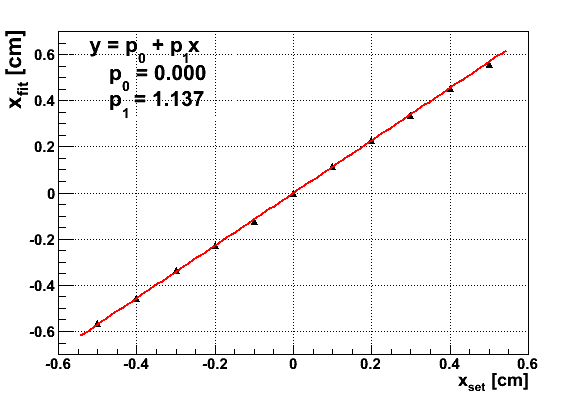}
\includegraphics[width=0.32\textwidth]{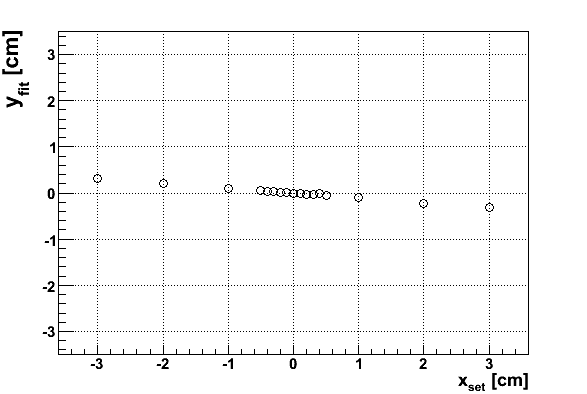}
\includegraphics[width=0.32\textwidth]{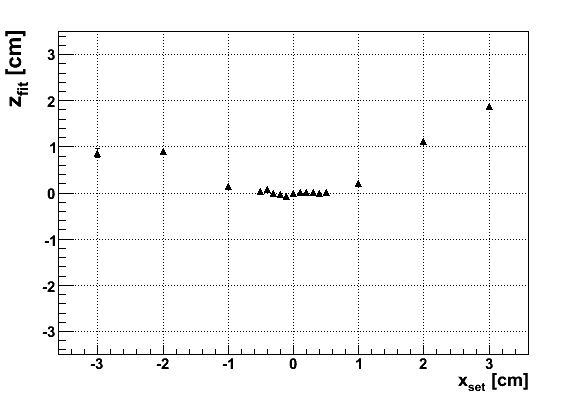}}
\centerline{%
\includegraphics[width=0.32\textwidth]{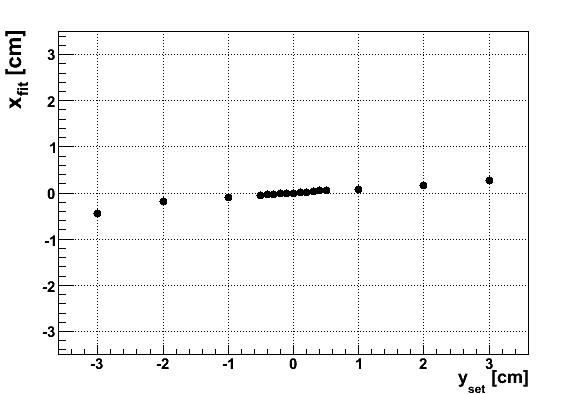}
\includegraphics[width=0.32\textwidth]{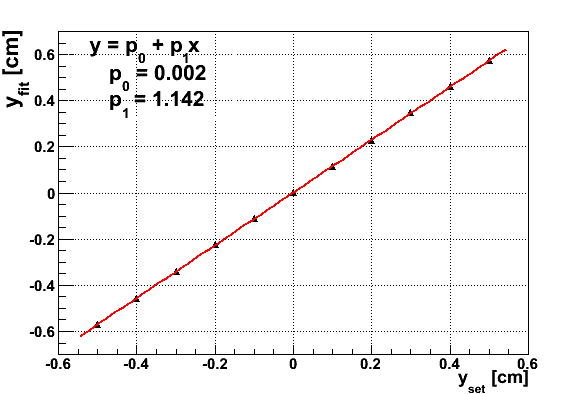}
\includegraphics[width=0.32\textwidth]{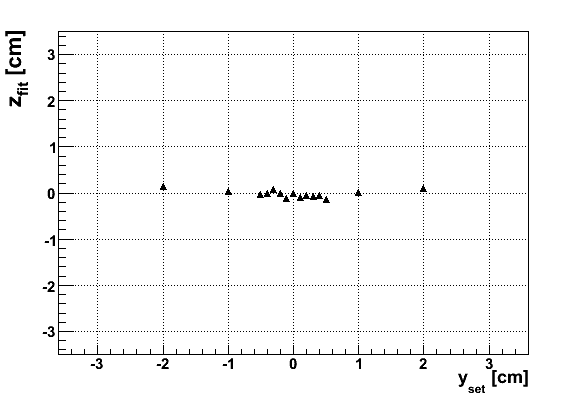}
}
\centerline{%
\includegraphics[width=0.32\textwidth]{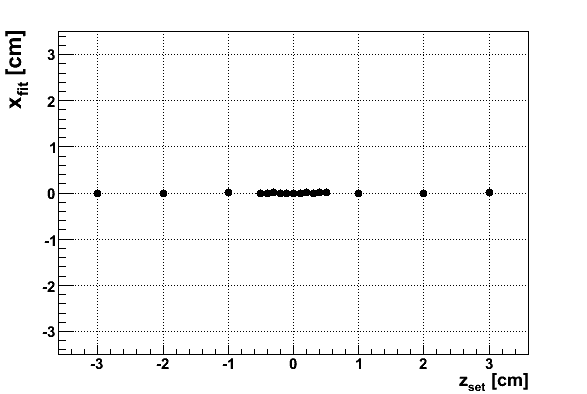}
\includegraphics[width=0.32\textwidth]{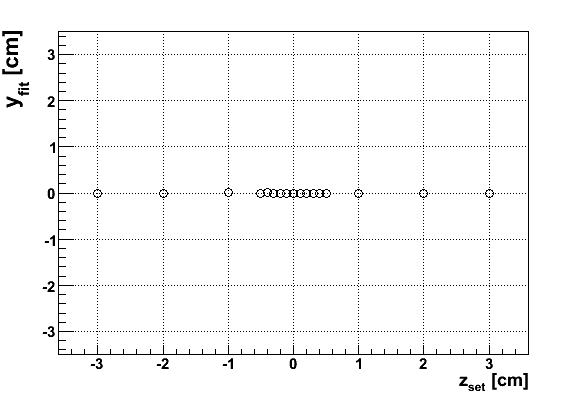}
\includegraphics[width=0.32\textwidth]{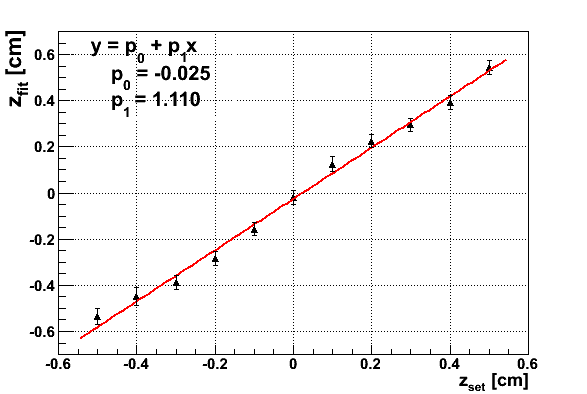}
}
\caption{Plots showing results of Monte Carlo tests made on the $x,y,z$~-coordinates
           of the reaction vertex (first, second and third row subsequently).  
           See text for details.}
\label{f:xyz_set}
\end{figure}

A set of simulations of elastic $pp$ scattering have been made with different locations of the vertex 
where only one of the vertex coordinates was changed at once, leaving the others at zero. The accuracy of the method
used to extract the vertex position \cite{Demirors2005} is shown in Fig.~\ref{f:xyz_set}.

In the first row, the plots corresponding to the change in the x-coordinate ($x_{set}$) of the vertex are shown.
In the second row the y-coordinate ($y_{set}$) was changed and in the third row the z-coordinate ($z_{set}$).
All plots are distributions of the extracted (fit) value of the given coordinate as a function
of the true value (set) of the coordinate being changed. Therefore, points on plots placed diagonally 
should be arranged along $fit(set) = set$ line while other distributions should show $fit(set) = 0$ behavior. 

The fits of the first order polynomial to the points on plots placed diagonally (red lines) show that
in all cases, the extracted values deviate slightly from the set ones (up to $14\%$ in case of the $y_{fit}(y_{set})$).
This need to be taken into account while extracting the vertex position in experimental data.
We can also notice, that if the change in a given coordinate is not bigger than about $0.5$~cm, the extraction 
of the other coordinates is accurate.

To determine how the wrong assumption about the vertex position affects the polarization, the polarization 
was calculated individually for each data sample, simulated 
with a change in the position of a certain coordinate.
Then, each of the simulated data samples was analyzed with the default assumption that the particle going
forward origins from the $(x_{v},y_{v},z_{v}) = (0,0,0)$ point \footnote{Tracking algorithm of the Mini Drift
Chamber do not assume a certain vertex position}. 

\begin{figure}[htb]
\centerline{%
\includegraphics[width=7cm]{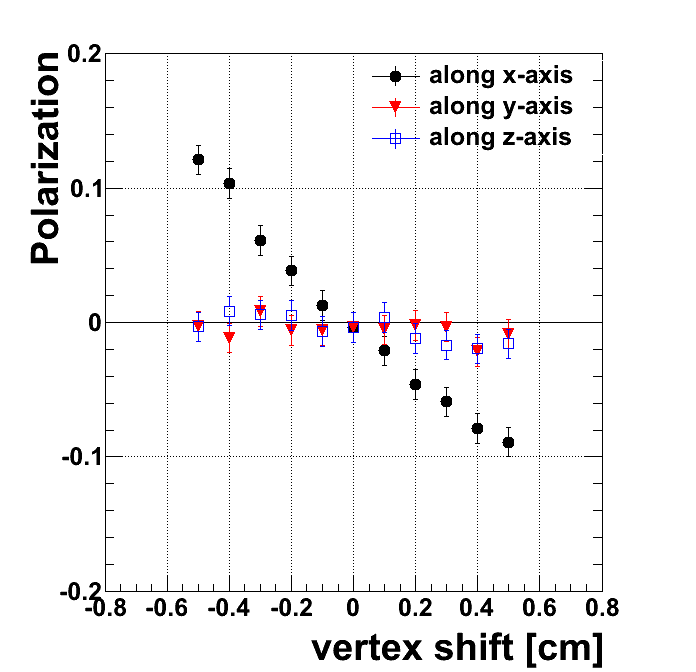}
\includegraphics[width=7cm]{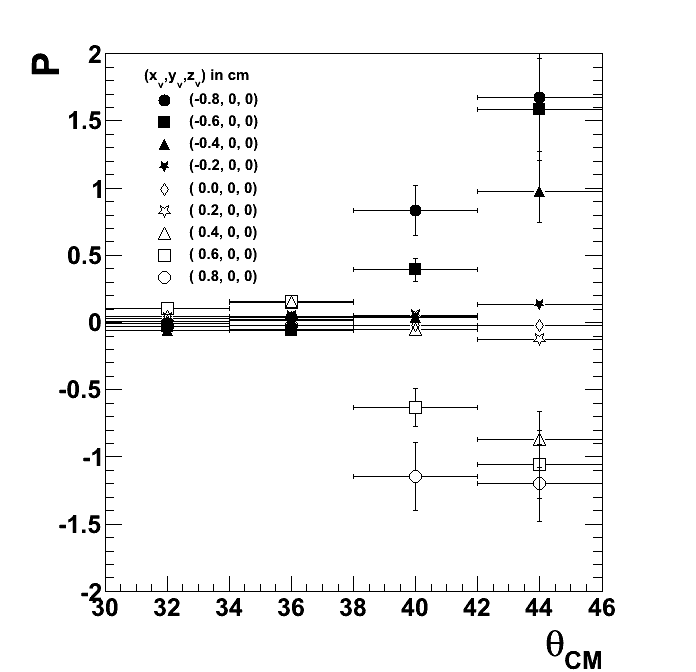}
}
\caption{Left: polarization vs. vertex shift along the $x-$, $y-$ and $z-$ axis (see the legend) 
determined assuming acceptance for the vertex position at $(0,0,0)$. Data were simulated at positions as indicated in the figure.
			Right: polarization as a function of the scattering angle of the forward 
					going proton (center-of-mass scattering), determined from the simulated data with
 different values of the $x-$coordinate of the interaction point (see the legend).  }
\label{f:pol_th_acc}
\end{figure}

The result is presented on the left panel of Fig.~\ref{f:pol_th_acc} which shows the determined polarization for different
vertex locations.

\begin{figure}
 \begin{minipage}[t]{.53\linewidth}\vspace{-2pt}\raggedright
  \begin{minipage}[t]{\linewidth}\vspace{0pt}\raggedright
   \includegraphics[width=\linewidth]{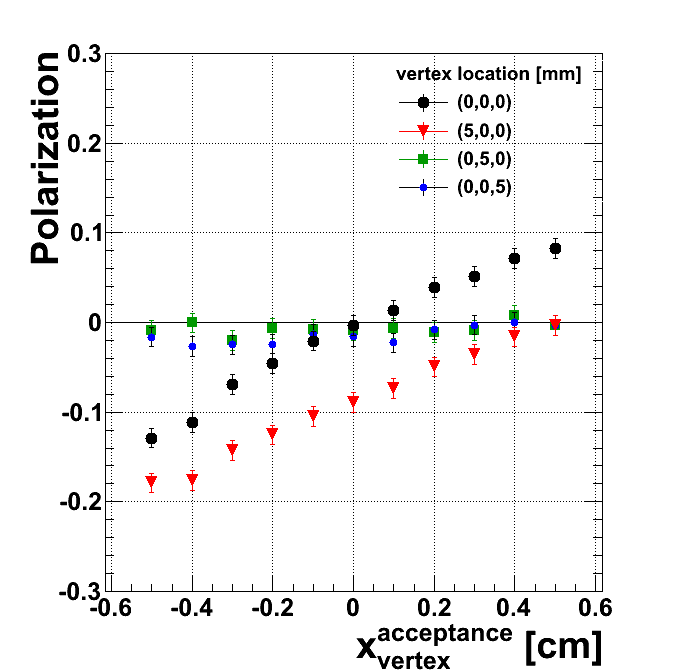}
  \end{minipage}
  \begin{minipage}[t]{0.9\linewidth}\vspace{40pt}\raggedright
\caption{Upper left panel: polarization vs. shift of the vertex location along the $x-$axis, taken in determining the acceptance correction, $x_{vertex}^{acceptance}$. The polarization was calculated for four locations of the interaction point (see the legend).
			Right side: coplanarity dependence on the protons' azimuthal angle.
}
\label{f:pol_cop}
  \end{minipage}
 \end{minipage}
 \hfill
 \begin{minipage}[t]{.48\linewidth}\vspace{0pt}
  \includegraphics[width=\linewidth]{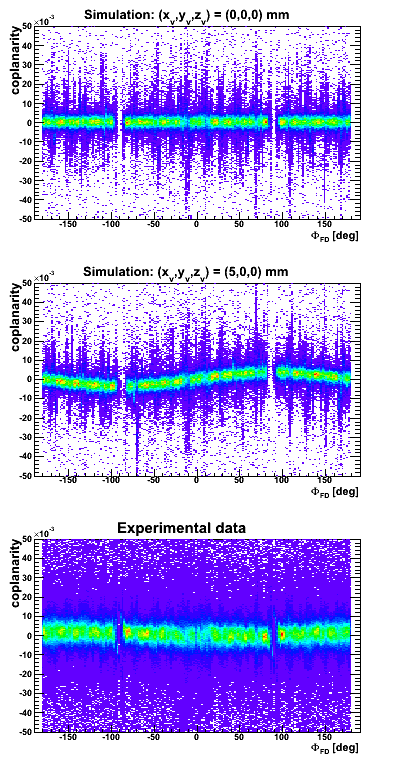}
 \end{minipage}

\end{figure}
While the change of the
$y_{v}$ or $z_{v}$ coordinate does not have influence on the result, a certain sensitivity of polarization is seen
in case of changing the $x_{v}$ coordinate of the interaction point. Namely, the calculated value of the polarization changes 
linearly with the shift of the vertex along the $x-$axis. The influence of moving the interaction point along the $x-$axis 
on the polarization depends on the scattering angle of the forward going proton, $\theta_{CMs}$. 
The right panel of Fig.~\ref{f:pol_th_acc} shows the distribution of the polarization as a function
of the scattering angle of the forward
going proton calculated in the centre of mass system, $\theta_{CMs}$,
made for different vertex positions (x-coordinate of the vertex was varied). It is seen, that for $\theta_{CMs} > 38^\circ$,
the polarization strongly deviates from the expected value when changing the $x_{v}$ coordinate by more than 5 mm.
Therefore, since the polarization for higher angles is biased 
by the systematics, we should restrict the used $\theta_{CMs}$ angle to less than $38^\circ$. On the other hand, the observed
dependency, if seen in experimental data, would be a clear sign of the wrong assumption of the $x-$position of
the interaction point. It is important to notice, that based on the results shown in Fig.~\ref{f:pol_cop} (left) 
the vertex position must be controlled with the accuracy better than $1$~mm in order to achieve uncertainties of the polarization determination of about  $0.03$.

On the upper, left panel of Fig.~\ref{f:pol_cop} a result of further studies is shown, how wrongly assumed location of the interaction point,
influences the polarization. Data simulated with four different vertex positions (as indicated in the legend), have been acceptance corrected
assuming different values of the $x_{v}$ coordinate, $x_{vertex}^{acceptance}$. 
In this case a result is similar as shown in Fig.~\ref{f:pol_th_acc}.  It shows that in order to control polarization determination with the precision of about $0.03$ we need to control the determination of the $x$-coordinate of the vertex with the precision of about $1$~mm. 
Comparison of black circles and red triangles indicates that this conclusion is independent of the 'true' position of the vertex, at least within the range of $5$mm. 
It might be noticed as well that data, generated with $y_{v}$ or $z_{v}$ set to $5$~mm and corrected to different $x_{vertex}^{acceptance}$, do not influence the polarization significantly.

Another way to control the location of the vertex position in the experiment is to monitor the coplanarity, $C$,
defined as
\begin{equation}
C = \frac{( \vec{p}_{1}\times \vec{p}_{2}) \cdot \vec{p}_{beam}}{ |\vec{p}_{1}\times \vec{p}_{2}| \cdot |\vec{p}_{beam}|},
\end{equation}
where $\vec{p}_{1}$ and $\vec{p}_{2}$ correspond to two scattered protons and $\vec{p}_{beam}$ is the vector of the beam.
The coplanarity dependence on the protons' azimuthal angle shows sinusoidal behavior for a misallocated vertex.
This is shown on the right side of Fig.~\ref{f:pol_cop}. The upper plot corresponds to simulated data with the vertex
located at $(x_{v},y_{v},z_{v}) = (0,0,0)$~cm and the $C(\phi)$ distribution is flat. Moving the vertex position
to point $(x_{v},y_{v},z_{v}) = (0.5,0,0)$~cm results in a sinusoidal shape. This is shown in the middle plot on the 
right side of Fig.~\ref{f:pol_cop}. Experimental data are presented in the lower, right corner of Fig.~\ref{f:pol_cop}.
Since, in this case, the events originate from all posibble vertex positions distributed within the region of the beam 
and target overlap, the points on the histogram should be arranged uniformly around $C(\phi)=0$.

\subsection{Tilt of the beam}
\begin{figure}[!htb]
\centerline{%
\includegraphics[width=6cm]{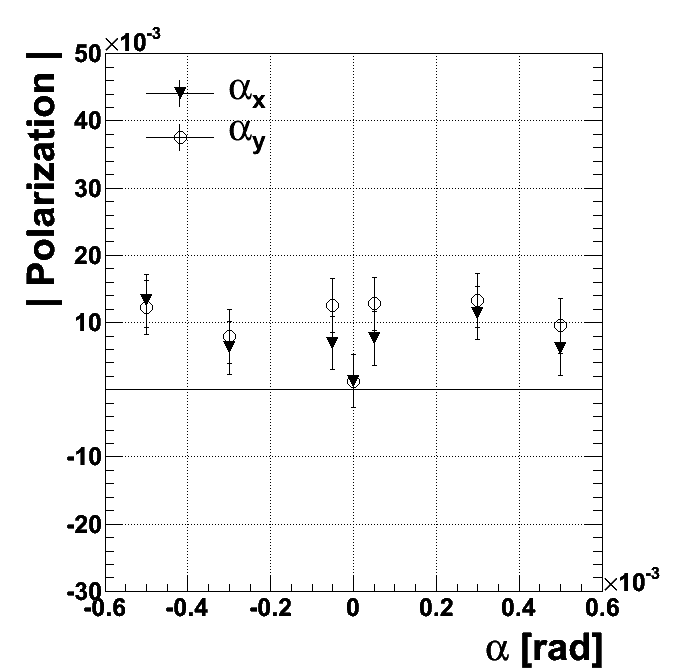}}
\caption{Distribution of the polarization as a function of the degree of the beam tilt in the $yz$-plane (filled triangles)
				and $xz$-plane (open circles). From simulations. The studied range is by factor of ten larger than the range of the possible tilt allowed by the COSY optics \cite{Prashun2013}.}
\label{f:th_ph_geant}
\end{figure}

The maximum allowed range of tilts of the beam at WASA-at-COSY is between $-0.05$~mrad and $0.05$~mrad (symmetrically
around the $z-$axis) \cite{Prashun2013}. To determine how the tilt of the beam affects the polarization,
the beam was leaned in the $yz$-plane or $xz$-plane at different $\alpha_{x}$ and $\alpha_{y}$ angles respectively.
In Fig.~\ref{f:th_ph_geant}, the polarization as a function of the $\alpha$ angle for both types of studied
beam tilts is shown. 
There are no effects observed in the studied range of the $\alpha$ angle ($\alpha \in [-0.5,0.5]$ mrad)
except that the polarization slightly differs from zero (up to $0.01$).

\subsection{Summary}
Methods to monitor the location of the vertex have been demonstrated and 
it was shown how a misallocation of the vertex impacts the obtained value
for the polarization. The study concluded that to have systematic uncertainty of the polarization smaller than
$0.03$, we need to control the position of the interaction point with a precision better than $1$~mm.
In this article we presented three methods for the determination of the vertex position:  
(i) based on the $d(\phi)$ distribution, 
(ii) coplanarity distribution, 
(iii) polarization as a function of $\theta_{CMs}$. 
Due to the large statistics of collected data, and the usage of the listed methods, the vertex position 
will be determined with a precision much better than $1$~mm. 
Due to the high sensitivity of the result to the scattering angle it is better to calculate the polarization
taking into account only scattering angles not bigger than $\theta_{CMs}=38^\circ$.

It was also presented that the beam, tilted within the maximum allowed range should
have no significant influence on the obtained degree values for the polarization.

\subsection{Acknowledgements}
We acknowledge support 
by the Polish National Science Center through grant No. 2011/03/B/ST2/01847, 
by the FFE grants of the Research Center Juelich, 
by the EU Integrated Infrastructure Initiative HadronPhysics Project under contract number RII3-CT-2004-506078 
and by the European Commission under the 7th Framework Programme through the ’Research Infrastructures’ action of the ’Capacities’ Programme, Call: FP7-INFRASTRUCTURES-2008-1, Grant Agreement N. 227431.


\begin{thebibliography}{99}  
\bibitem{Abdel-Bary2003} 
 M. Abdel-Bary et al., Eur. Phys. J. A16 (2003) 127
\bibitem{Moskal2004}
 P. Moskal et al., Phys. Rev. C 69 (2004) 025203
\bibitem{Petren2010}
 H. Petren et al., Phys. Rev. C 82 (2010) 055206
\bibitem{Moskal2010}
 P. Moskal et al., Eur. Phys. J. A43 (2010) 131
\bibitem{Calen1996}
 H. Calen et al., Phys. Lett. B366 (1996) 39-43
\bibitem{Calen1997}
 H. Calen et al., Phys. Rev. Lett. 79 (1997) 2642-2645
\bibitem{Hibou1998}
 F. Hibou et al., Phys. Lett. B438 (1998) 41-46
\bibitem{Smyrski2000}
 J. Smyrski et al., Phys. Lett. B474 (2000) 182-187
\bibitem{Bergdolt1993}
 A. M. Bergdolt et al., Phys. Rev. D48 (1993) 2969-2973
\bibitem{Moskal2009}
 P. Moskal at al., Phys. Rev. C79 (2009) 015208
\bibitem{Moskal2004b}
 P. Moskal, hep-ph/0408162 (2004)
\bibitem{AbdEl2001}
 S. Abd El-Samad et al., Phys. Lett. B522 (2001) 16-21 
\bibitem{Czyz2007}
 R. Czyzykiewicz et al., Phys.Rev.Lett. 98 (2007) 122003
\bibitem{Winter2003}
 P. Winter et al., Eur. Phys. J. A18 (2003) 355
\bibitem{Balestra2004}
 F. Balestra et al. Phys. Rev. C69 (2004) 064003
\bibitem{MoskalHodana2010}
 P. Moskal, M. Hodana, J. Phys. Conf. Ser. 295 (2011) 012080
\bibitem{Altmeier2000}
 M. Altmeier et al. Phys.Rev.Lett. 85 (2000) 1819-1822 
\bibitem{Demirors2005}
 L. Demirors PhD Hamburg Univerity (2005)
\bibitem{Prashun2013}
 D. Prashun private communication (2013)



\end{thebibliography}
\end{document}